\documentclass[aps,prl,reprint,superscriptaddress,showpacs]{revtex4-1}
\usepackage{amsmath,amssymb,bm}
\usepackage{graphicx}
\usepackage{epstopdf}
\usepackage{latexsym}
\usepackage{subfigure}
\usepackage{color}
\usepackage{hyperref}

\newcommand{\ccc}{{Cs$_2$CuCl$_4$}}
\newcommand{\be}{\begin{equation}}
\newcommand{\ee}{\end{equation}}
\newcommand{\bea}{\begin{eqnarray}}
\newcommand{\eea}{\end{eqnarray}}

\begin{document}

\title{Magnetic resonance in the ordered phases  of \\ the 2D frustrated quantum magnet Cs$_2$CuCl$_4$}

\author{A. I. Smirnov}
    \affiliation{P. L. Kapitza Institute for Physical Problems, RAS, 119334 Moscow, Russia}
    \affiliation{Moscow Institute for Physics and Technology, 141700, Dolgoprudny, Russia}

\author{K. Yu. Povarov}
        \affiliation{P. L. Kapitza Institute for Physical Problems, RAS, 119334 Moscow, Russia}
    \author{S. V. Petrov}
    \affiliation{P. L. Kapitza Institute for Physical Problems, RAS, 119334 Moscow, Russia}
\author{A. Ya. Shapiro}
    \affiliation{A. V. Shubnikov Institute of Crystallography, RAS, 119333 Moscow, Russia}

\date{\today}

\begin{abstract}
The temperature evolution of the electron spin resonance is studied at cooling the crystal samples of \ccc\
through the N\'{e}el point 0.62 K. A coexistence of the high-frequency spinon type resonance developed in the
spin-liquid phase and of the low-frequency antiferromagnetic resonance was found in the ordered phase. The
low-frequency magnetic resonance spectrum in the low field range has two gapped branches and corresponds well to
the spectrum of spin excitations of a planar spiral spin structure with two axes of the anisotropy. The field
induced phase transitions result in a more complicated low-frequency spectra.
\end{abstract}


\maketitle

\section{Introduction \label{Introduction}}

\begin{figure}[h]
\includegraphics[width=10pc]{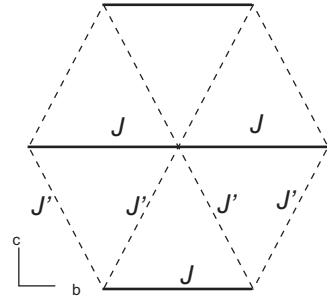}\hspace{2pc}%
\caption{\label{fig:JandJprime} Sketch of the exchange paths of \ccc \ within the $bc$-plane }
\end{figure}

Magnetic crystals with low dimensional and frustrated antiferromagnetic exchange interaction remain in a
correlated paramagnetic state (often called "spin liquid phase") far below the Curie-Weiss temperature $T_{CW}$.
The suppression or even absence of the ordering is due to strong quantum fluctuations. A quasi 2D $S=1/2$
dielectric antiferromagnet \ccc \ is a thoroughly studied example of a magnetic system with spin correlations
emergent in a wide temperature range below $T_{CW}$=4~K, but still above the ordering point $T_N$=0.62 K. In
crystals of \ccc \ the Cu$^{2+}$ magnetic ions are placed on stacked 2D layers with distorted triangular
lattice. Experiments on inelastic neutron scattering in the spin-liquid phase uncovered magnetic excitations
propagating along $b$-axis with  a spectrum of an extensive multiparticle continuum \cite{ColdeaPRL,ColdeaPRB}.
The wide band of transferred energy corresponding to  a fixed transferred momentum $q$ is observed for $q
\backsim \frac{1}{b}$, where $b$ is a lattice period, while at low momenta $q\ll \frac{1}{b}$ the spectrum width
is negligible in zero magnetic field. The boundaries and the spectral density of the continuum observed in \ccc
\ correspond well to the two-spinon continuum of a quantum critical  spin S=1/2 chain
\cite{ColdeaPRL,KohnoStarykhBalents}. The 1D nature of spinon excitations in this 2D structure is attributed to
the frustration of exchange bonds on the distorted triangular lattice, with the strong exchange bonds ($J$=0.375
meV) directed along $b$-direction, and a weaker exchange interaction ($J^\prime=0.34J$), coupling magnetic
Cu$^{2+}$ ions along lateral sides of isosceles triangles, see Fig. \ref{fig:JandJprime}. The interlayer
exchange $J^{\prime\prime}$=0.045$J$ is the weakest one. The spin chains extended along $b$-direction should be
practically decoupled from each other due to the geometric frustration of the exchange bonds $J^\prime$, as
shown in numerical simulations \cite{Becca} and in the analytical approach \cite{KohnoStarykhBalents}. Another
consequence of the frustration is the extreme sensitivity of the ground state to the weak interlayer exchange
and Dzyaloshinsky-Moriya interaction \cite{Starykh2010}. The study of elastic neutron scattering
\cite{ColdeaPRL} showed, that the ordering at $T_N$ results in a spiral spin structure. The spiral lies
approximately in the $bc$-plane and its wavevector is directed along $b$-axes. The ordered spin component has a
strong quantum reduction $\Delta S/S$=0.25 \cite{ColdeaJPCM}.

\begin{figure}[h]
\includegraphics[width=16pc]{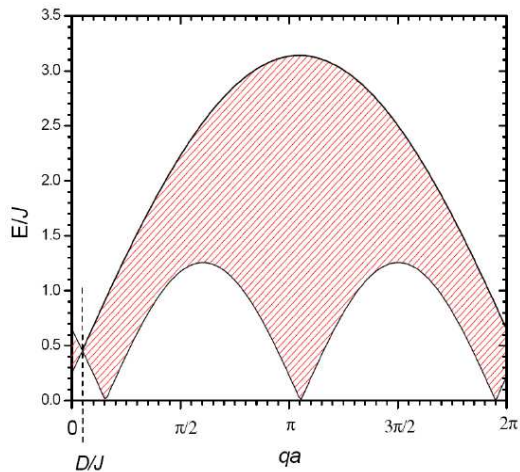}\hspace{2pc}%
\begin{minipage}[b]{16pc}\caption{\label{continuum} The continuum of transversal spin excitations for 1D spin chain
with the uniform Dzyaloshinsky-Moriya interaction in a magnetic field $H=0.5J/(g\mu_B)$ at $\bf{H}
\parallel \bf{D}$. Note a considerable width of the continuum at $q$=0. }
\end{minipage}
\end{figure}

The spinon continuum in \ccc \ corresponds mainly to that of S=1/2 Heisenberg antiferromagnetic chain.
Nevertheless, it  is modified in the low-frequency range by the uniform Dzyaloshinsky-Moriya interaction of
magnetic ions within a chain \cite{StarykhESR, povarov,affleck2011}. The uniform Dzyaloshinsky-Moriya
interaction between the spins in a chain is a distinct feature of \ccc. The corresponding model Hamiltonian is
derived in Ref. \cite{Starykh2010},  it's terms $D_a$ and  $D_c$ present this interaction. Besides, the
Dzyaloshinsky-Moriya interaction of a conventional, i.e. staggered, type is present for the ions connected by
exchange bonds $J^{\prime}$. This interaction corresponds to  terms $D_{a}^{\prime}$ and $D_{c}^{\prime}$ of the
Hamiltonian of Ref. \cite{Starykh2010}. The uniform interaction results in the shift of the continuum spectrum
$E(q)$ along the $q$ axis for a value of $\delta q$=$\sqrt{D_a^2+D_c^2}/(Jb)$. A zero field gap arises due to
this shift.  The shift results also in an essential change of the envelope of the spectrum of spin fluctuations
at $q$=0 in a magnetic field aligned parallel to the Dzyaloshinsky-Moriya vector ${\bf D}=(D_a,D_c)$.  The $q=0$
spectrum, which in the absence of uniform Dzyaloshinsky-Moriya interaction is represented by a single Zeeman
frequency $\mu_BH/\hbar$,  transforms in a band of frequencies corresponding to the width of the continuum,
$\sim J\delta q$, see Fig. 2. This results in an effective splitting of the electron spin resonance (ESR) line
at $\bf{H}
\parallel \bf{D}$, because the spectral density has singularities at the boundaries of the continuum
\cite{StarykhESR}. For $\bf{H} \perp \bf{D}$, there is no splitting, while the gap in the ESR spectrum causes an
essential shift of the ESR line. The  zero-field gap  in the ESR spectrum, shift of the resonance field at ${\bf
H}
\parallel b$ and split doublet at ${\bf H} \perp b$, in accordance with vector $\bf {D}$ orientation in \ccc\ were indeed
observed \cite{povarov} in the temperature range of the spin-liquid phase. The gap at $T=2T_N=$1.3 K has a value
of 14 GHz, about a half of the zero-temperature gap in the ordered phase.

The aim of the present work is to study the temperature evolution of the above exotic spinon-type  ESR (with the
gap opening above $T_N$ and anisotropic splitting) at cooling through $T_N$, as well as the study of the
antiferromagnetic resonance at low temperatures. We observe a surprising ESR spectrum at low temperatures: for
the higher frequency band,  above 60 GHz, the spinon-type doublet ESR was observed to be preserved till the
lowest temperature, while for low frequencies, below 40 GHz the spinon magnetic resonance is replaced by the
antiferromagnetic resonance.
The coexistance of spinon resonance and antiferromagnetic resonance at low temperatures presents a new kind of
ESR spectrum of a spin S=1/2 quantum antiferromagnet.

\begin{figure}[h]
\includegraphics[width=\columnwidth]{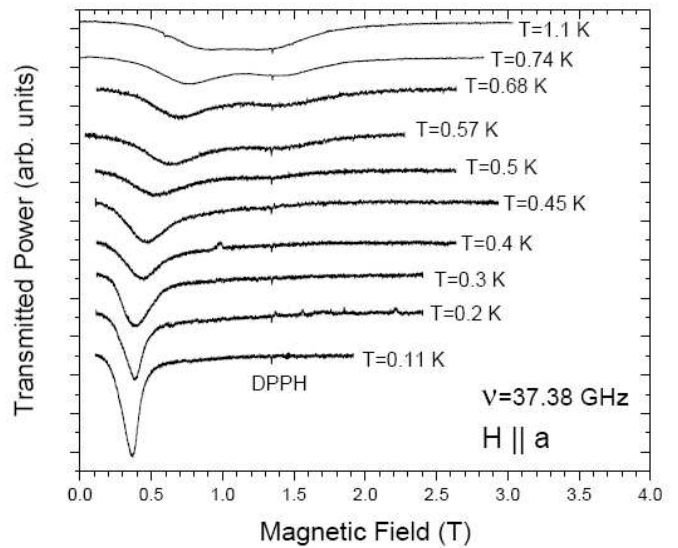}
\caption{\label{lowfreq} Temperature evolution of the  ESR signal in the low frequency range at ${\bf
H}\parallel a$. Weak sharp resonances at $H$=1.35 T is a g=2.0 label of free radicals in
diphenyl-picryl-hydrazil (DPPH).}
\end{figure}

\begin{figure}[h]
\includegraphics[width= \columnwidth]{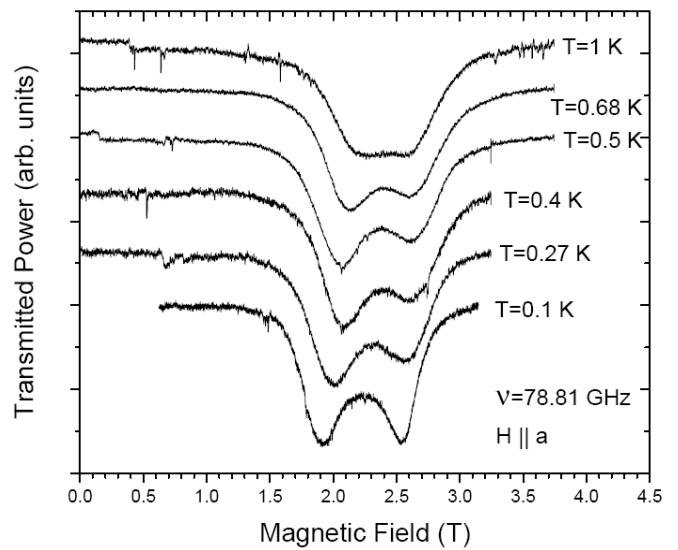}
\caption{\label{highfreq}Temperature evolution of the  ESR signal in a high frequency range at ${\bf H}\parallel
a$.}
\end{figure}

\begin{figure}[h]
\includegraphics[width=\columnwidth]{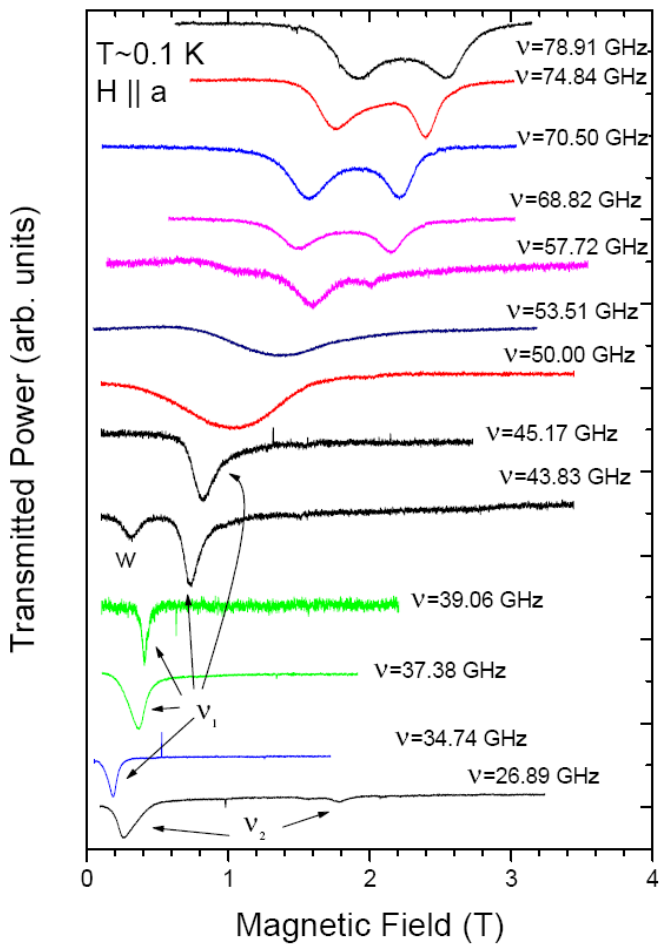}\hspace{2pc}%
\caption{\label{HIIaLinesvarf} ESR lines  at $T<$0.1 K and ${\bf H}\parallel a$ for different frequencies}
\end{figure}

\begin{figure}[h]
\includegraphics[width=\columnwidth]{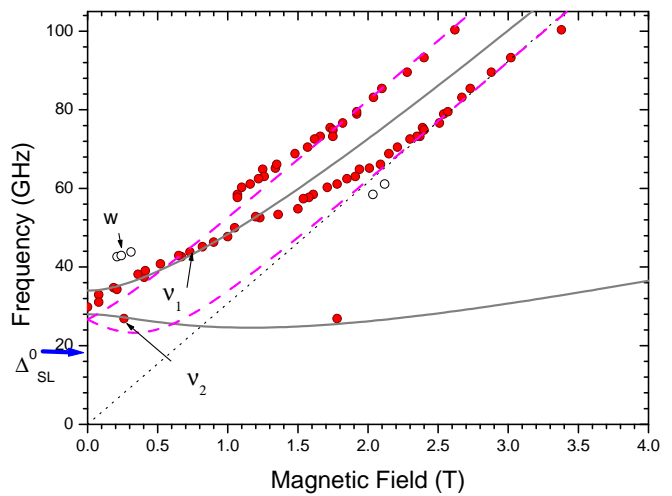}\hspace{2pc}%
\caption{\label{fvshaLT} Frequency-field dependence for ESR in \ccc \  at $T<$0.1 K and ${\bf H}\parallel a$.
Filled symbols correspond to the intensive ESR modes, empty symbols  present resonances of a weak intensity.
Solid lines correspond to the calculation according to eq. (\ref{Aspektr}). Dashed lines present the proposed
spinon-type resonances, $\nu_{\pm}$, according to relation (\ref{formulaNupm}). Dotted line corresponds to the
ESR of noninteracting spins with $g=g_a$.}
\end{figure}

\section{Experiment}

The magnetic resonance signals were recorded as field dependences of the transmitted microwave power in the
frequency range 25$<$$\nu$$<$140 GHz, using multimode microwave resonator and waveguide insert combined with a
dilution refrigerator Kelvinox-400. The sample was fixed by the Apeizon N grease inside the copper resonator,
placed in vacuum and connected via a thermoconducting link to the mixing chamber of the dilution refrigerator.
The temperature of the resonator was monitored by a RuO$_2$ thermometer placed on the top of the resonator.  To
avoid heating of the resonator unit during experiments at the temperatures below 1 K,  the screening of the
infrared thermal radiation, passing through the waveguides, was performed by two filters thermalized at 80 K and
at 4 K, following  Ref. \cite{ULTESR}. The incident microwave power at the cryostat input was not higher than 1
$\mu$W. Microwave absorption by the sample is estimated as 1 nW, for this estimation we used the measurements of
the transmission of the waveguides with filters and of the coupling between the resonator and waveguides, as
well as the typical ratio of microwave losses in the sample and in the resonator walls.  The overheating of the
sample due to microwave absorption is estimated as 10 mK at $T=$0.1 K. Considering the above estimations we can
not surely distinct the temperatures of the sample below 0.1 K, despite the observation that thermometer at the
top of the resonator indicated temperature down to 0.05K during the field sweep. We checked the sample heating
effects by making experiments with different values of the incident microwave power. At enlarging the power we
marked the power level causing overheating of the resonator or change of the ESR lineshape or position. After
that, an attenuation for more than a factor of 2 was inserted for the incident power. The samples from the same
batch as in Ref. \cite{povarov} were used. As reported in Ref. \cite{povarov}, the ESR signal at $T>$10 K is a
typical single mode resonance corresponding to $g$-factor values $g_{a,b,c}$=2.20, 2.08, 2.30 for the
orientation of the magnetic field along the crystallographic axes $a$, $b$ and $c$ correspondingly. At cooling
below 4 K a significant change of ESR lineshape and position was found: for the temperature $T$=1.3 K and  ${\bf
H}
\parallel b$ the resonance line is shifted to lower fields, while for ${\bf H} \parallel a,c$ it is transformed to a
doublet due to the modification of the spinon continuum described above. Now we describe the transformation of
the ESR lines at further cooling through the ordering point $T_N$=0.62 K.

\subsection {Field along $a$-axis}
At first, we consider the temperature evolution of ESR signals for the most simple case of ${\bf H}\parallel a$.
Here the  magnetic field is perpendicular to the spin spiral plane and the structure exhibits just gradual
transformation of the cone configuration to fully polarized state without intermediate phase transitions
\cite{ColdeaPRL,Tokiwa}.  We observed two kinds of the temperature evolution of the doublet ESR signals at
cooling below $T_N$: i) At low frequencies ($\nu<$ 40 GHz), the spinon doublet is gradually transformed into a
single  narrow line of the antiferromagnetic resonance, as shown in Fig. \ref{lowfreq}. ii) For higher
frequencies, $\nu>$ 60 GHz, the doublet found in the spin-liquid phase survives deep in the ordered phase, see
Fig. \ref{highfreq}. In the intermediate frequency range 40$<$$\nu$$<$60 GHz,  there is a single ESR line, which
is, however, strongly broadened with respect to observations in the low frequency range. Examples of the low
temperature ESR lines at different frequencies are displayed on Fig. \ref{HIIaLinesvarf}.  The ESR lines, which
arise via the temperature evolution of the first type, are marked here as $\nu_1$. Besides, there is a
low-frequency branch of the ESR (marked as $\nu_2$ on Figs. \ref{HIIaLinesvarf}, \ref{fvshaLT}), which was
observed at the frequency 26.9 GHz and only in the ordered phase. The collection of resonance fields observed at
$T<$0.1 K in the whole frequency range is presented on the frequency-field diagram Fig. \ref{fvshaLT}. Weak
resonances, marked by "w" on Figs. \ref{HIIaLinesvarf}, \ref{fvshaLT} were observed at several frequencies in
the intermediate frequency range 40-60 GHz. In summary, for ${\bf H}
\parallel a$,  at $T<$0.1 K  we observe in the low frequency range narrow ESR lines corresponding to two
branches - $\nu_1$,  with a rising frequency-field dependence,  and a low frequency branch $\nu_2$. In the high
frequency range above 60 GHz there is a doublet of ESR lines analogous to that observed in the spin-liquid
phase.

\begin{figure}[h]
\includegraphics[width=\columnwidth]{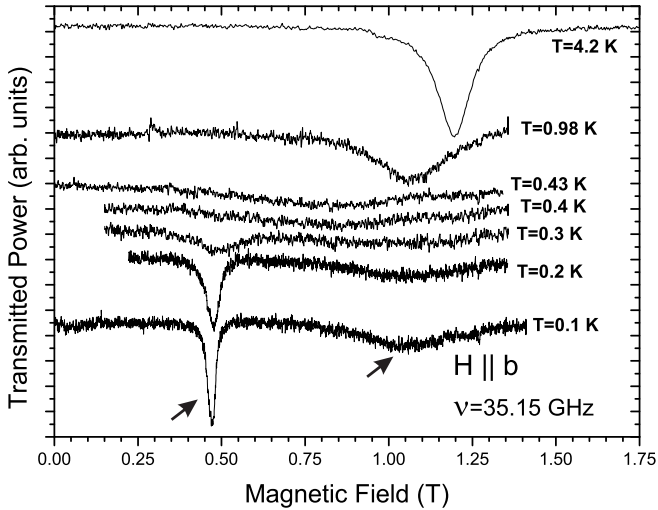}
\caption{\label{HIIbEvolT} Temperature evolution of the 35.15 GHz ESR signal at ${\bf H}\parallel b$.}
\end{figure}

\begin{figure}[h]
\includegraphics[width=\columnwidth]{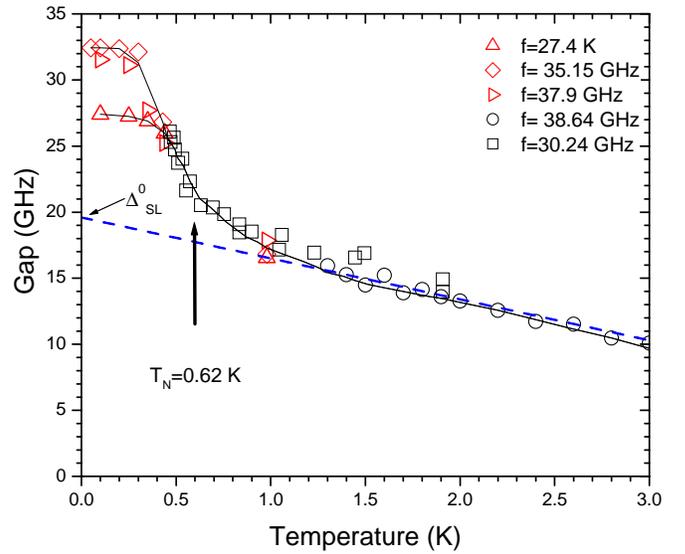}
\caption{\label{gapvsT} Temperature dependence of the gap $\Delta$, calculated from  the equation
(\ref{formulaGap}) using the resonance field values at different frequencies for ${\bf H} \parallel b$ in the
spin-liquid and ordered phases. Solid lines are guide to the eyes. Dashed line is the extrapolation of the
spin-liquid energy gap to $T=0$.}
\end{figure}

\begin{figure}[h]
\includegraphics[width=\columnwidth]{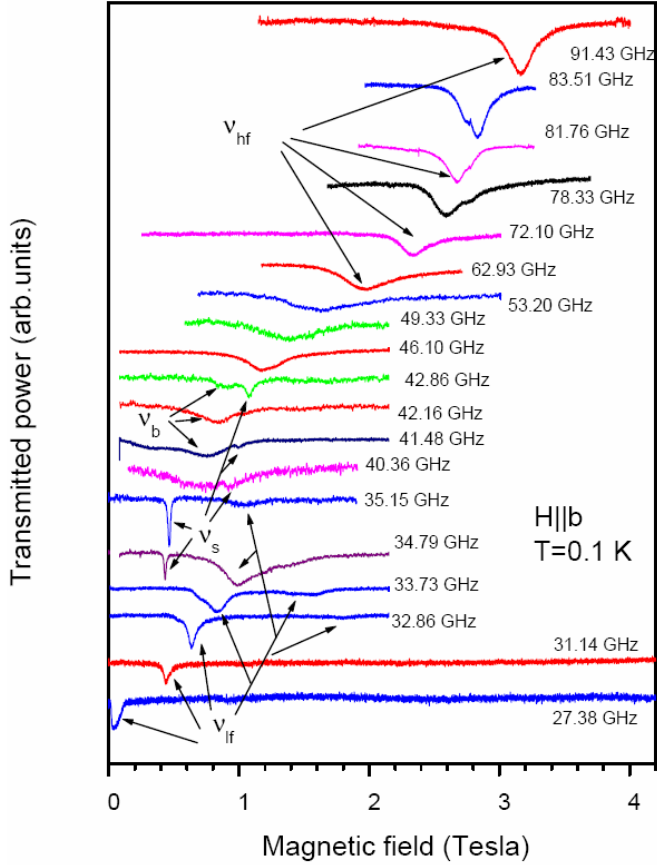}
\caption{\label{HIIbLinesvarf} Examples of low temperature ESR lines at ${\bf H} \parallel b$, taken at
different frequencies.}
\end{figure}

\begin{figure}
\includegraphics[width=\columnwidth]{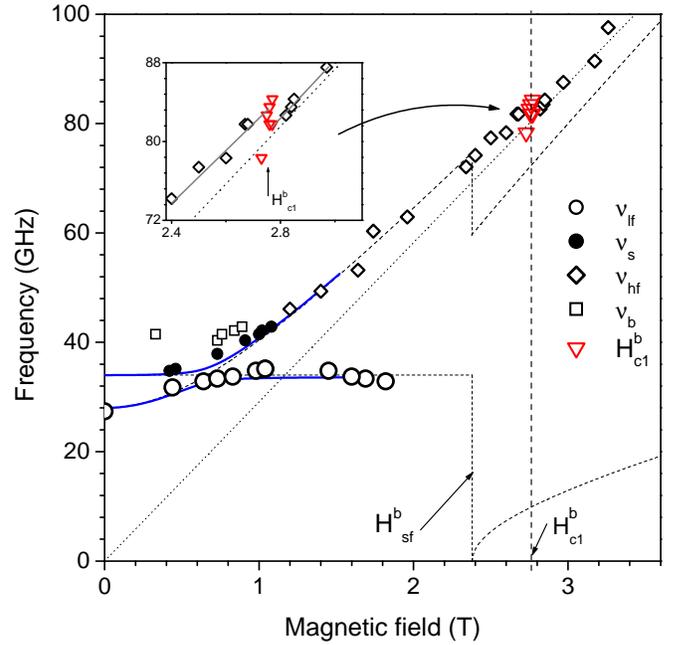}
\caption{\label{fvshb}Frequency-field diagram for T=0.1 K and ${\bf H} \parallel$ b. Triangles mark the field
$H^{b}_{c1}$ of the phase transition, measured on ESR records, other symbols represent resonances, marked in
Fig.\ref{HIIbLinesvarf}. Dashed lines present the calculation according to the macroscopic theory for ${\bf
H}\parallel b$, solid lines - for 5$^\circ$ tilting angle between $\bf{H}$ and $b$. Dotted line is the
paramagnetic resonance frequency for $g$=2.08. Field $H^{b}_{sf}$ represents calculation by formula
(\ref{Bflop}) of the Appendix . Vertical dashed line represent the field $H^{b}_{c1}$ measured in
Ref.\cite{Tokiwa}. Insert: frequency-field dependence in the vicinity of the phase transition.}
\end{figure}

\begin{figure}[h]
\includegraphics[width=0.7\columnwidth]{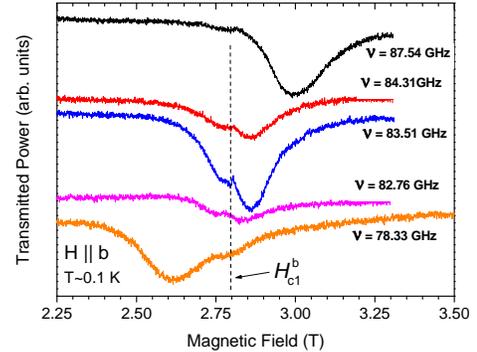}
\caption{\label{HIIbTrans} ESR signals near the field-induced phase transiton at $H^b_{c1}=$2.75 T, ${\bf
 H}\parallel b$.} \hspace{2pc}
\end{figure}

\subsection {Field along $b$-axis.}

 For ${\bf H} \parallel b$,  there is a single ESR line in the spin-liquid phase. This resonance is shifting
 to lower fields with cooling. The shift of the resonance field from the magnetic field of the ESR
 of noninteracting spins $H_0=2\pi \hbar \nu /(g_b \mu_B)$ becomes visible at temperatures below 4 K.
On further cooling, below  the N\'{e}el temperature, the resonance line is strongly broadened near $T_N$ and
than reappears as a narrow line at $T\sim$0.3 K, this evolution is illustrated in Fig.\ref{HIIbEvolT}. Another
resonance mode appears at low temperatures at higher field. These two modes are marked by arrows on Fig.
\ref{HIIbEvolT}. In the temperature range of the spin-liquid state the value of the shifted resonance field may
be described by the relation for a conventional gapped resonance (see Ref. \cite{povarov}):

\begin{equation}
 \nu^2=(\frac{g_b\mu_B H}{2 \pi \hbar})^2+ \Delta^2
 \label{formulaGap}
\end{equation}

The temperature dependence of the gap $\Delta$, derived from the experiments on different frequencies, using the
relation (\ref{formulaGap}) is presented on Fig. \ref{gapvsT}. The gap is gradually rising on cooling from $T=$
4K with a steep increase below $T_N$. In the ordered phase, below 0.4 K,  we resolve two modes with different
gaps, the resonance fields of both modes were used for  calculation of the $\Delta(T)$ dependence presented on
Fig.\ref{gapvsT}. The low-temperature limit of the lowest gap 28 GHz is twice as large as the gap at $T_N$. The
low-field resonance on Fig. \ref{HIIbEvolT} corresponds to a branch with a rising frequency {\it vs} field
dependence and has a narrow linewidth $\Delta H_s \sim$ 0.02 T in the frequency range 35-43 GHz. These
resonances are marked as $\nu_s$ on Fig.\ref{HIIbLinesvarf}, where examples of ESR lines taken at different
frequencies are presented.  The resonances, observed at low temperature are collected on the frequency-field
diagram Fig. \ref{fvshb}. A family of low-frequency resonances with the frequencies below 35.5 GHz, including
the upper field resonance of Fig. \ref{HIIbEvolT} is marked as $\nu_{lf}$ on Figs. \ref{HIIbLinesvarf},
\ref{fvshb}. At higher frequencies, 40-45 GHz, the resonance $\nu_s$ is superimposed with a broad line $\nu_b$
with a width of $\Delta H_b \sim$ 0.5 T. In the range 45-50 GHz the broad line $\nu_b$ appears alone, and at
higher frequencies the narrower line $\nu_{hf}$ with a width $\Delta H_{hf} \sim$ 0.2 T is observed. Thus, the
spectrum undergoes a crossover between the low frequency sharp resonance mode $\nu_s$ and the high frequency
resonance $\nu_{hf}$ via coexisting resonances and a broad resonance in the intermediate frequency range
40$<\nu<$60 GHz.

For the fields in the vicinity of 2.8 T we observe an abrupt modification of the ESR lines, which is clearly
seen on Figs. \ref{HIIbLinesvarf}, \ref{HIIbTrans}. This modification corresponds to a small step in the
$\nu(H)$ dependence, presented on the insert of Fig. \ref{fvshb}. The field $H=2.8 \pm 0.05$ T of the step on
the $\nu(H)$ dependence corresponds well to the field-induced phase transition at $H^{b}_{c1}$=2.75 T observed
in Ref. \cite{Tokiwa} as a jump on the magnetization curve.

\subsection{Field along $c$-axis}

\begin{figure}[h]
\includegraphics[width=\columnwidth]{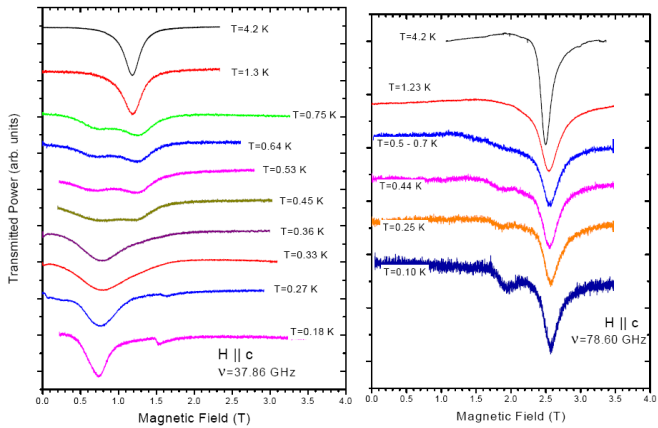}
\caption{\label{HIIcTeevolvarfreq} ESR signals at ${\bf H}\parallel c$ for different temperatures.} \hspace{2pc}
\end{figure}

\begin{figure}[h]
\includegraphics[width=0.7\columnwidth]{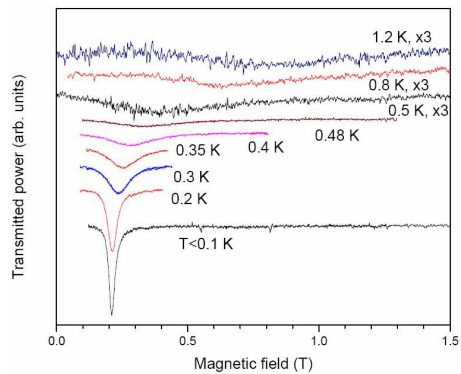}
\caption{\label{HIIcTeevol34p5Ghz} ESR signals at ${\bf H}\parallel c$ for different temperatures at 34.5 GHz.}
\hspace{2pc}
\end{figure}

\begin{figure}[h]
\begin{minipage}{18pc}
\includegraphics[width=\columnwidth]{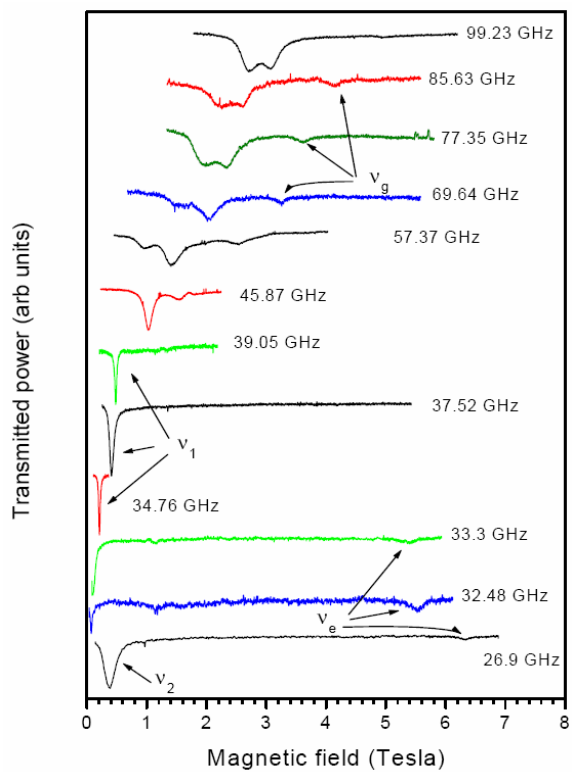}
\caption{\label{HIIc_lines} ESR signals at different frequencies for ${\bf H}\parallel c$, T$<$0.1 K}
\end{minipage}\hspace{2pc}
\end{figure}

\begin{figure}[h]
\begin{minipage}{18pc}
\includegraphics[width=\columnwidth]{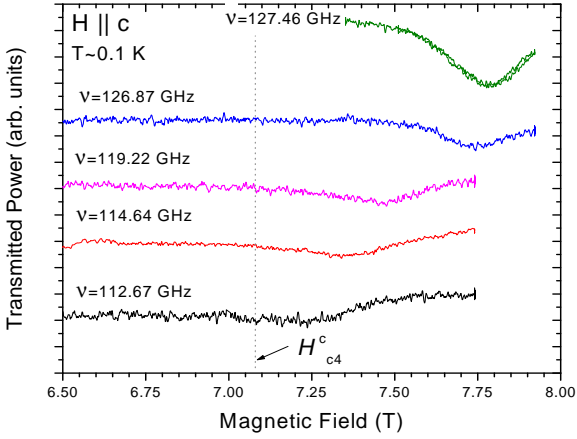}
\caption{\label{HFresonanceHIIc} ESR signals  of the high-field mode $\nu_h$. ${\bf H}\parallel c$, $T<$0.1 K.}
\end{minipage}\hspace{2pc}
\end{figure}

\begin{figure}[h]
\begin{minipage}{18pc}
\includegraphics[width=\columnwidth]{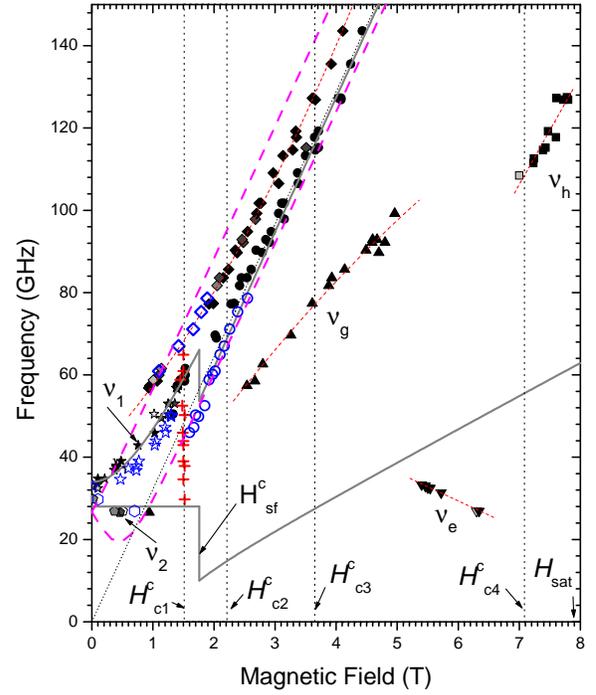}
\caption{\label{HIIc_fvsH} frequency-field dependence for ESR signals at ${\bf H}\parallel c$. Filled  and empty
symbols  correspond to  two different samples, crosses - to values of $H^c_{c1}$, measured as kinks on ESR
records at different frequencies. Solid grey lines present macroscopic theory, thin dashed lines are guide to
eyes, thick dashed lines present the proposed spinon-type resonances $\nu_{\pm}$, according to relation
(\ref{formulaNupm}). Dotted line presents the high-temperature ESR. Vertical dashed lines correspond to critical
fields $H^c_{c1,2,3,4}$, observed in magnetization experiments in Ref. \cite{Tokiwa}. The saturation field
$H_{sat}$ is marked according to ref. \cite{Tokiwa}. The field $H^c_{sf}$ presents the calculated spin-flop
field according to formula (\ref{Cflop}).}
\end{minipage}\hspace{2pc}
\end{figure}

\begin{figure}[h]
\begin{minipage}{18pc}
\includegraphics[width=0.7\columnwidth]{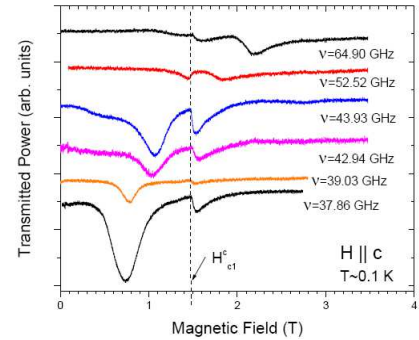}
\caption{\label{HIIcTransition} ESR signals at ${\bf H} \parallel c$ near the field induced transition.
$H^c_{c1}$ is marked according to Ref.\cite{Tokiwa}.}
\end{minipage}\hspace{2pc}
\end{figure}

For ${\bf H} \parallel c$, a  doublet of ESR lines appears in the spin liquid phase. At further cooling, as in
the case of ${\bf H} \parallel a$,  for $\nu<$ 40 GHz we observe the evolution to a single ESR line, while for
$\nu>$ 60 GHz the doublet survives at low temperatures, as presented in Fig. \ref{HIIcTeevolvarfreq}. Fig.
\ref{HIIcTeevol34p5Ghz} presents the temperature evolution of the resonance line till the lowest temperature
$T<$0.1 K, demonstrating the evolution to the narrow line with the width $\Delta H$=0.04T. The resonance lines
at different frequencies are given in Figs. \ref{HIIc_lines}, \ref{HFresonanceHIIc}, the frequency-field diagram
is presented on Fig. \ref{HIIc_fvsH}. At low temperature, in the low-field range we observe a strongly rising
resonance branch, and a slowly falling one, marked as $\nu_1$ and $\nu_2$ correspondingly on Figs.
\ref{HIIc_lines}, \ref{HIIc_fvsH}. In the low field range the $\nu_1$ branch is observed as very narrow ($\Delta
H_1$=0.03-0.04 T) resonance, while in the frequency range above 50 GHz (corresponding to the field range 1.5-2.5
T) the resonance line is distorted, and at frequencies higher than 70 GHz there is again a doublet, analogous to
the case of ${\bf H}\parallel a$. In the frequency range 37-64 GHz the  ESR records  have pronounced kink, which
marks the value of the field of the phase transition at $H^c_{c1}=1.48$T, as illustrated on Fig.
\ref{HIIcTransition}. The step on the frequency-field dependence at $H^c_{c1}$ corresponds well to the phase
transition observed by magnetization measurement at the same field as a step in magnetization and a jump in the
susceptibility \cite{Tokiwa}. Besides, three low-frequency branches of lower intensities are detected in higher
fields. Two of these branches are marked as $\nu_{e,g}$ on Figs. \ref{HIIc_lines}, \ref{HIIc_fvsH}. The third,
high-field resonance mode, $\nu_h$, is presented on Fig. \ref{HFresonanceHIIc}, its frequency-field dependence
is shown on Fig. \ref{HIIc_fvsH}. The resonances $\nu_{e,g}$ constitute a branch emerging above the field of the
phase transition at $H^c_{c2}=2.1$ T, also observed in Ref. \cite{Tokiwa}. The branch $\nu_h$ is observable in
the field range between 7 and 8 T. This field range corresponds to a high-field phase, which was observed in
Ref. \cite{Tokiwa} between the fields $H^c_{c4}$=7 T and $H_{sat}$=8 T.

\section{Discussion}

The nature of the spin ordering in  \ccc \ is a curious subject, because of the frustration of the lateral
exchange bonds, causing the decoupling of spin chains. The ordered phases and field-induced transitions of \ccc
\ are discussed in detail in Ref. \cite{Starykh2010}. At $H$=0, the ordering is supposed to be due to the joint
action of the in-chain exchange $J$, the Dzyaloshinsky-Moriya interaction between ions on lateral bonds of
triangular lattice ($D_a^{\prime}$-term of Ref. \cite{Starykh2010}) and the frustrated exchange $J^{\prime}$.
This interplay should cause the "DM-spiral" with the spins lying in the $bc$-plane (see Ref.
\cite{Starykh2010}), as confirmed by the experiment \cite{ColdeaPRL}. At the magnetic field ${\bf H} \parallel
a$ the spiral ordering of transversal spin components takes place until the saturation field, as expected in
theory and observed in experiment. However, at ${\bf H} \parallel b, c$ and $H\simeq D_a^\prime$ the spin spiral
is expected  to flop perpendicular to the magnetic field and the above $D_a^{\prime}$-term becomes ineffective
for the flopped configuration. At ${\bf H} \parallel b$,  the phase should be than ordered due to the action of
$J$ and $J^{\prime\prime}$, resulting in the collinear AF phase.    At ${\bf H} \parallel c$, another
Dzyaloshinsky-Moriya term, $D_c^\prime$ becomes effective in the flopped phase and the most complicated sequence
of phase transitions occurs. The spin structures for field-induced phases at ${\bf H} \parallel b, c$ are
proposed in Ref. \cite{Starykh2010} by the consideration of the interactions characterized by the parameters $J,
J^\prime, J^{\prime \prime}, D^{\prime}_{a,c} $. The uniform Dzyaloshinsky-Moriya interaction, parametrised by
$D_{a,c}$ terms does not contribute to the selection of the long range ordered ground states because its sign
oscillates from one chain to another.  According to the analysis  of ordered phases \cite{Starykh2010}, for
${\bf H}
\parallel b$, two transitions before saturation are predicted, and for ${\bf H} \parallel c$ - four transitions
before saturation. This  corresponds well to the observations of Refs. \cite{ColdeaPRL,Tokiwa,Tokigawa}, where
the field-induced transitions are observed for ${\bf H} \parallel b$  in fields $H^b_{c1,2}$,  and for  ${\bf H}
\parallel c$ in fields $H^c_{c1,2,3,4}$. These fields, except for $H^b_{c2}$ are marked on Figs. \ref{fvshb},
\ref{HIIc_fvsH}.

 We interpret magnetic resonance in the ordered phase of \ccc \  in low fields, where the planar spin spiral
is stable for all directions of the magnetic field.  The antiferromagnetic resonance of a planar spiral
structure with two axes of the anisotropy should obey two gapped modes of the antiferromagnetic resonance and a
zero-frequency Goldstone mode, see, e.g. Ref. \cite{SvistovFarutin}. Two modes with different gaps correspond to
out of plane oscillation of the spiral spin system and the third mode with a zero frequency represents in-plane
rotation of the spiral which may be performed without the energy loss. The low-frequency dynamics of the planar
spiral antiferromagnet was described in Ref. \cite{SvistovFarutin} using the macroscopic approach developed in
Ref. \cite{AndreevMarchenko}. In this approach one considers the spiral spin structure parametrised by two
orthogonal unit vectors $\mathbf{l}_{1}$ and $\mathbf{l}_{2}$:

\begin{equation}
S=\mathbf{l}_{1}\sin({\bf qr})+\mathbf{l}_{2} \cos({\bf qr})
\end{equation}

  The consideration is valid in the exchange approximation, i.e. in the magnetic fields far below the
saturation ($H_{sat}\simeq$ 8 T  for \ccc ) and for low frequency $2\pi\hbar\nu \ll J$, it does not involve the
resonance absorption due to the transitions of magnons over a gap at the wavevector of the spiral. The resonance
frequencies obtained by this approach are represented in Sec. Appendix, equations (\ref{Bspektr}-\ref{Aspektr}).
This consideration predicts also spin-flop transitions for ${\bf H}\parallel b,c$ in the magnetic fields
$H^{b,c}_{sf}$ given by (\ref{Bflop},\ref{Cflop}). At these fields the spin plane  should flop perpendicular to
the magnetic field because of the anisotropy of the susceptibility.

The experimental frequencies for all three directions of the magnetic field were fitted by the relations
(\ref{Bspektr}-\ref{Aspektr}) in the low field range. The exchange approximation does not involve the anisotropy
of $g$-factor. We include the anisotropy of $g$-factor by taking the value of $\gamma=g_{a,b,c}\mu_B/\hbar$ for
orientations of $\bf{H}$ along the axes $a,b,c$ correspondingly. The frequency-field dependences, calculated for
${\bf H}\parallel a,c$ are presented  on Figs. \ref{fvshaLT}, \ref{HIIc_fvsH} by solid lines. For the
orientation ${\bf H}\parallel b$ the intersection of calculated branches takes place. This causes a repulsion of
branches in case of a tilting of the magnetic field with respect to the $b$-axis. The frequency, calculated for
a perfect ${\bf H}\parallel b$ orientation is presented on Fig. \ref{fvshb} by dashed line, while $\nu(H)$,
modeled for the tilting angle of 5$^{\circ}$ is given by a solid line. Calculated curves for both perfect and
tilted configurations agree with the experimental resonance frequencies, while the  calculation for the tilted
field gives a better coincidence with the evolution of the resonance field at a gradual increase of the
frequency.

The fitting parameters are $\omega_{10}/(2\pi)=34 \pm 2$ GHz and $\omega_{20}/(2\pi)=28 \pm 2$ GHz,
$A=-7.2\cdot10^{6} \text{ erg/mole}=-8.6\cdot10^{-2}\text{ K/Cu}$ and $B=-2.3 \text{
erg/mole}=-2.7\cdot10^{-3}\text{ K/Cu}$. The energy of the anisotropy, presented  by constants $A$ and $B$ are
much smaller than the main exchange integral $J=4.3\text{ K}$ and the constant of uniform Dzyaloshinsky-Moriya
interaction derived in Ref. \cite{povarov} $D=0.4\text{ K}$. The constant of the staggered Dzyaloshinsky-Moriya
interaction, derived from neutron scattering experiments in  polarised state is $D_a^\prime$=0.23 K.  Parameter
$\eta$ was calculated using the susceptibility values measured in Ref.\cite{Tokiwa}:
$\chi_{\parallel}=4.9\cdot10^{-2}$ and $\chi_{\perp}=3.7\cdot10^{-2}$ emu/mole for $T\rightarrow 0$. It is worth
to note that the parameter $\eta$, calculated in such a way, does not correspond exactly to the parameter of the
anisotropy of the susceptibility considered in the theory of Ref. \cite{SvistovFarutin}, because the anisotropy
of the susceptibility is of the same order as that of the $g$-factor, while the theoretical model considers
parameter $\eta$ in the exchange approximation with an isotropic $g$-factor. Besides, the condition of the low
frequency approximation $\hbar\omega \ll J$ is not valid for the case of \ccc, because observed gaps (~30 GHz)
are about a half of $J/(2\pi\hbar)$.
 Due to violation of conditions of applicability and above cited approximations, we consider the
 results given in the Appendix
as providing only a qualitative description and classification of modes of antiferromagnetic resonance and
giving the approximate values of spin-flop transitions for \ccc. In particular, the absence of a good
correspondence between the calculated spin flop fields $H_{sf}^{b,c}$ and the observed values $H^{b,c}_{c1}$ is
attributed to these approximations. The values of observed \cite{Tokiwa} field-induced transitions and of the
calculated spin-flop fields are marked on Figs. \ref{fvshb},\ref{HIIc_fvsH}.

Thus, the  experiment in the lower range, $\nu <$ 40 GHz, corresponds qualitatively to the macroscopic approach
(\ref{Bspektr}-\ref{Aspektr}). However, increasing the frequency to a value of the order of the exchange
frequency $J/(2\pi\hbar)$=67 GHz and higher, we observe at ${\bf H}\parallel a, c$ a gradual transformation of
the upper branch of the antiferromagnetic resonance $\nu_1$  into a doublet of lines. A transformation of ESR
line is also observed for the third orientation, ${\bf H}\parallel b$, there is an evolution of a narrow ESR
line to a broadened one via an intermediate resonance at 40 GHz $< \nu <$ 50 GHz. Therefore, the spectrum
observed in the high frequency range, is analogous to that in the spin-liquid phase, i.e. a doublet for ${\bf
H}\parallel a,c$ and a single gapped line for ${\bf H}\parallel b$.

To check this, we suggest to extrapolate the hypothetic spinon-type doublet ESR frequencies $\nu_{\pm}$  with
the spin-liquid ansatz based on equations (4,5) of Ref. \cite{povarov}. Specifically, we attempt to describe
resonance frequencies of the doublet with the relation

\begin{equation}
 \nu_{\pm}=\nu_{R,L}+\delta \nu,
 \label{formulaNupm}
\end{equation}

where  $\nu_{R,L}$ are taken from (4,5)  of Ref. \cite{povarov} but with extrapolated values  $D_a=$ 10.4 GHz
and $D_c=$ 14.3 GHz which have been increased by 30\% from their measured at $T=$ 1.3 K  values. This correction
should be made because the gap induced by Dzyaloshinsky-Moriya interaction, $\Delta=\frac{\pi}{2}
\sqrt{D_{a}^2+D_{c}^2}$, measured  in the spin liquid phase above $T_N$ is observed to slow  increase with
lowering of the temperature, see Fig. \ref{gapvsT}. This suggests, that the zero-temperature values of $D_{a}$
and $D_{c}$ are larger, than values extracted from  $T=1.3$ K measurements. Of course, these parameters cannot
be measured directly at low temperature, where the inter-chain exchange and other residual interactions
establish the three-dimensional long range order. The ordering produces a steep increase of the gap in the
neighborhood of $T_N$=0.62 K, see Fig.\ref{gapvsT}. Nevertheless, we can estimate the $T=0$  "spin-liquid gap"
extrapolating the temperature dependence of $\Delta$ in Fig. \ref{gapvsT} from the spin-liquid range. Linear
extrapolation of $\Delta(T)$, shown by the dashed line on Fig. \ref{gapvsT}, results in $\Delta^0_{SL}$=19.6
GHz, and represents a 30 \% increase of $D_{a,c}$  from their  $T=1.3$ K values when $\Delta$=15 GHz. The second
term in eq. (\ref{formulaNupm}),  $\delta \nu =\frac{\omega_{20}}{2\pi}-\Delta^{0}_{SL}$=9 GHz is an empiric way
to account for the renormalization of the gap by the three dimensional magnetic ordering. The values of
$\nu_{\pm}$ calculated in this way are shown by thick dashed lines on Figs. \ref{fvshaLT}, \ref{HIIc_fvsH}.   We
see that the high-frequency experimental data correspond well to this quasi-empirical proposition for ${\bf
H}\parallel a$ and there is a qualitative correspondence for ${\bf H}\parallel c$, i.e. the splitting of the
doublet observed in the ordered phase has a natural value for the spin liquid state extrapolated to zero
temperature. Thereby, the changes in the ESR lineshape and spectrum observed at increasing the frequency for all
three principal orientations ${\bf H}
\parallel a,b,c$  mark a crossover from antiferromagntic resonance to a spinon type ESR. In addition,
following the temperature evolution of ESR (Figs. \ref{highfreq}, \ref{HIIcTeevolvarfreq} ) from the spin liquid
to the ordered phase we also can interpret the doublet observed in the high frequency range $\nu>$ 60 GHz at
${\bf H}\parallel a, c$ as the spinon type ESR formed in the spin-liquid phase and surviving deep in the ordered
phase.

Thus, if the magnetic field is large enough, i.e. $\mu_B H \simeq J$, we may conclude, that the two spinon
continuum near $q\simeq$0 remains unchanged in the ordering process and further cooling. At lower fields (and
lower frequencies $\nu < J/(2\pi\hbar)$), for $q$=0, the continuum is replaced by antiferromagnetic resonance of
the ordered spiral structure.  The spectrum with the coexistance of the spinon type ESR and of the
antiferromagnetic resonance indicates  a new kind of  magnetic resonance, arising due to the specific ground
state with  the ordered spin components and strong zero point fluctuations. As far we know, there is no adequate
theory for this type of ESR.

The observation  of the coexistance of two kinds of ESR signals in the ordered phase of \ccc \ is naturally
related to a similar feature of the inelastic neutron scattering experiments of Ref. \cite{ColdeaPRB}. In these
(zero-field) experiments, at cooling below $T_N$, the spinon continuum was found to remain practically unchanged
for the energy range above 0.2 meV (i.e 50 GHz), while in the narrow interval near the lowest boundary a spin
wave mode at ~0.1 meV (25 GHz) was found below $T_N$, see Fig. 6 of Ref. \cite{ColdeaPRB}. In our experiments we
observe that the spinon-type ESR remains approximately undistorted at high frequency, while it is replaced by
the antiferromagnetic resonance at low frequency.  We can not still interpret the weak modes $\nu_{e,g,h}$,
observed for  field-induced phases at $\bf{H} \parallel c$ because the spin structure for these phases is formed
under a complex  interplay of several interactions.

\section{Conclusion}

   The electron spin resonance, corresponding to a two spinon continuum of quantum critical $S=1/2$ spin chain
   was found to exist both above and below the ordering point $T_N$
   in the frequency range above the exchange frequency $J/(2\pi\hbar)$. At the same time,
   in the low frequency range, the spinon type ESR is replaced by  the
   antiferromagnetic resonance below the N\`{e}el point.

\section{Acknowledgements}
Authors acknowledge K. O. Keshishev for guidance in use of the dilution refrigerator and O. A. Starykh,
S.S.~Sosin, for useful discussions. The work was supported by Russian Foundation for Basic Research, grants
11-02-12225 and 12-02-00557.

\section{Appendix}

The Lagrange function of a mole of a planar spiral magnet is

\begin{eqnarray}
    \mathcal{L} &=& \frac{\chi_{\parallel}(1-\eta)}{4\gamma^{2}}( (\dot{\mathbf{l}}_{1}+
     \gamma [\mathbf{l}_{1}\times \mathbf{H}])^{2}  + \nonumber\\
 && (\dot{\mathbf{l}}_{2}+\gamma [\mathbf{l}_{1}\times \mathbf{H}])^{2} ) \nonumber\\
    &&+\frac{\chi_{\parallel}(1+\eta)}{4\gamma^{2}}(\dot{\mathbf{n}}+\gamma [\mathbf{n}\times
    \mathbf{H}])^{2}-U_{a}
    \label{FarutinLagr}
\end{eqnarray}

Here $\mathbf{n}=[\mathbf{l}_{1}\times\mathbf{l}_{2}]$ and   $\eta=1-\dfrac{\chi_{\perp}}{\chi_{\parallel}}$,
$\chi_\parallel$ is the susceptibility along $\mathbf{n}$, and $\chi_\perp$ - is the susceptibility in the spin
plane. $\gamma=\mu_B/\hbar$ is the giromagnetic ratio for the spin angular momentum.  The energy of magnetic
anisotropy may be taken in the quadratric form as

\begin{equation}
    U_{a}=\frac{1}{2}(An_{z}^{2}+Bn_{y}^{2})
    \label{FarutinAnisotr}
\end{equation}

Taking the axes as follows $a\rightarrow z || \mathbf{n}$, $b\rightarrow x$, $c\rightarrow y$, with $A<0$,
$A<B$, we have in the ground state a spiral within the $bc$ plane. Varying the Lagrangian (\ref{FarutinLagr})
with respect to  $\delta \mathbf{l}_{1}$ è $\delta \mathbf{l}_{2}$ we get the resonance frequencies.

1) for ${\bf H} \parallel b$

\begin{equation}
\label{Bspektr} \omega=\omega_{10},\phantom{a}\sqrt{\omega_{20}^2+(\gamma H)^2}
\end{equation}

2) ${\bf H} \parallel c$

\begin{equation}
\label{Cspektr} \omega=\omega_{20},\phantom{a}\sqrt{\omega_{10}^2+(\gamma H)^2}
\end{equation}

3) ${\bf H} \parallel a$

\begin{multline}
\label{Aspektr} \omega^{2}=\frac{\omega_{10}^{2}+\omega_{20}^{2}}{2} +
(\gamma H)^{2}\frac{2+\eta^{2}}{4} \pm {}\\
{}\pm ( \frac{(\omega_{10}^{2}-\omega_{20}^{2})^{2}}{4} + 2(\gamma
H)^{2}(\omega_{10}^{2}+\omega_{20}^{2})\frac{(1+\eta)^{2}}{4}+ \\
4(\gamma H)^{4}\frac{(1-\eta^{2})^{2}}{16}
                             )^{1/2}
\end{multline}

Here $\omega_{10}^{2}=\dfrac{-A}{\chi_{||}}\gamma^{2}$, $\omega_{20}^{2}=\dfrac{B-A}{\chi_{||}}\gamma^{2}$.

For  $\chi_{\parallel}>\chi_{\perp}$  there should be a spin flop transition  at the in-plane magnetic fields
$H_{sf}$, at which the spin plane flops perpendicular to the magnetic field. These critical fields may be
calculated in the same low-field approximation for the magnetic field along $b$ and $c$ correspondingly:
\begin{equation}
    H_{sf}^{b}=\frac{\omega_{10}}{\gamma\sqrt{\eta}}
    \label{Bflop}
\end{equation}

\begin{equation}
    H_{sf}^{c}=\frac{\omega_{20}}{\gamma\sqrt{\eta}}
    \label{Cflop}
\end{equation}

 For  $H\parallel b$ the resonance frequencies at $H> H_{sf}^{b}$
should be calculated by formula (\ref{Aspektr}), with $\omega_{10}^{2}=\dfrac{A}{\chi_{\parallel}}\gamma^{2}$ è
$\omega_{20}^{2}=\dfrac{B}{\chi_{\parallel}}\gamma^{2}$.

For $H\parallel c$ at $H> H_{sf}^{c}$, the resonance frequencies should be calculated by (\ref{Aspektr}) with

$\omega_{10}^{2}=-\dfrac{B}{\chi_{\parallel}}\gamma^{2}$ and
$\omega_{20}^{2}=\dfrac{A-B}{\chi_{\parallel}}\gamma^{2}$.


\end{document}